\documentclass[11pt,oneside,a4paper]{article}
\usepackage{fullpage}
\usepackage[affil-it]{authblk}
\usepackage{amsmath}
\usepackage{amsthm}
\usepackage{amsfonts}
\usepackage{amssymb}
\usepackage{mathtools}
\usepackage{multirow}
\usepackage{url}
\usepackage{booktabs}
\usepackage{subcaption}
\usepackage{array}
\usepackage{hyperref}
\usepackage{adjustbox}
\usepackage{indentfirst}

\newcommand{\eg}{e.\,g., }
\newcommand{\ie}{i.\,e., }

\newcolumntype{H}{>{\setbox0=\hbox\bgroup}c<{\egroup}@{}}

\providecommand{\keywords}[1]
{
  \small	
  \textbf{\textit{Keywords---}} #1
}

\providecommand{\conflicts}[1]
{
  \small	
  \textbf{\textit{Conflicts of Interest:}} #1
}

\providecommand{\fund}[1]
{
  \small	
  \textbf{\textit{Funding:}} #1
}

\providecommand{\correspondence}[1]
{
  \small
  {\textbf{\textit{Correspondence to:}}}~#1
}

\providecommand{\data}[1]
{
  \small	
  \textbf{\textit{Data statement:}} #1
}


\title{\textbf{Metrics  to  Detect  Small-Scale  and  Large-Scale  Citation  Orchestration}}
\author{Iakovos Evdaimon\textsuperscript{\rm 1}, John P. A. Ioannidis\textsuperscript{\rm 2}, Giannis Nikolentzos\textsuperscript{\rm 3}, Michail Chatzianastasis\textsuperscript{\rm 1}, George Panagopoulos\textsuperscript{\rm 4}, and Michalis Vazirgiannis\textsuperscript{\rm 1, 5}\\ 
\textnormal{\textsuperscript{\rm 1}LIX, \'Ecole Polytechnique, Institut Polytechnique de Paris, Rte de Saclay, Palaiseau, 91120, France}\\
\textnormal{\textsuperscript{\rm 2}Meta-Research Innovation Center at Stanford (METRICS), and Departments of Medicine, of Epidemiology and Population Health, and of Biomedical Data Science, Stanford University, Stanford, CA 94305}\\
\textnormal{\textsuperscript{\rm 3}Department of Informatics and Telecommunications, University of Peloponnese, Akadimaikou G. K. Vlachou Street 221 31 Tripoli, Greece}\\
\textnormal{\textsuperscript{\rm 4}Department of Computer Science, University of Luxembourg, Maison du Nombre
6, avenue de la Fonte L-4364 Esch-sur-Alzette, Luxembourg}\\
\textnormal{\textsuperscript{\rm 5}Mohamed bin Zayed University of Artificial Intelligence, Masdar City, SE45 05, Abu Dhabi, United Arab Emirates}
}

\date{}

\begin{document}
\maketitle

\begin{abstract}
Authors may increase or inflate their citation metrics with legitimate or gaming means. Here, we explore citation orchestration phenomena. We define small-scale orchestration, when the author him/herself and/or a small number of other authors use citations strategically to boost citation metrics like $h$-index; and large-scale orchestration, when extensive collaborations among many co-authors lead to high h-index for many/all of them. 
We use Scopus data to investigate orchestration across science among $1,496,680$ authors with $\textgreater5$ full papers and $\geq1,000$ citations. We propose three orchestration indicators: extremely low ratio of citations over the square of the h-index $C/h^2$ (indicative of small-scale orchestration); extremely small number of authors who can explain at least $50\%$ of an author’s citations $A_{50\%C}$ (indicative of either small-scale or large-scale orchestration); and extremely large number of co-authors with $\textgreater50$ co-authored papers $A_{50}$ (indicative of large-scale orchestration). We explore a $1\%$ percentile threshold for these indicators as potential indicative of orchestration. Authors with extreme $C/h^2$ or $A_{50\%C}$ are typically prolific and have high h-index, while authors with extreme $A_{50}$ are typically highly prolific and have extremely high h-index. Orchestration indicators have diverse prevalence across scientific disciplines. Extremely low $C/h^2$ and $A_{50\%C}$ often co-exist, while extremely low $C/h^2$ rarely co-occurs with extremely high $A_{50}$. Using the Retraction Watch database, we document several authors with retractions among those with extreme orchestration indicators, especially with extremely low $C/h^2$. Overall, the proposed metrics do not prove misdeed, and each case should be examined separately, but they may offer insights into citation gaming practices.
\end{abstract}


\conflicts{None}

\data{Detailed data are available from the authors upon request.}

\correspondence{Iakovos Evdaimon, evdaimon@lix.polyetchnique.fr $|$ John P. A. Ioannidis, \indent jioannid@stanford.edu}

\fund{None}

\section{Introduction}
Citation counts and related metrics are widely used and misused in scholarly evaluation~\cite{bornmann2014excellent}.
They reflect the acknowledgment and utilization of research by the scientific community and play a significant role in assessing an author's research impact and contributions within their respective fields~\cite{costas2015altmetrics,ioannidis2014bibliometrics}.
However, the conventional popular metrics used to quantify research impact may fall short in capturing the intricate relationships between authors and their citation patterns~\cite{waltman2012inconsistency}.

For example, the $h$-index, introduced by Hirsch in 2005, has gained considerable popularity as a bibliometric indicator, measuring an author's productivity and citation impact~\cite{hirsch2005index}.
It represents the maximum number of an author's papers that have received at least $h$ citations.
While the $h$-index provides a straightforward measure of an author's research impact, it overlooks the underlying complexities of citation patterns and the potential manipulation of citation metrics~\cite{bornmann2005does}. Extensive criticism has been yielded against the $h$-index and its gaming potential~\cite{KELLY2006167,ca_chapman,ja_oravec}. 

In the scholarly landscape, there is an emerging concern regarding the strategic use of citations to enhance the influence of certain individuals or groups.
This phenomenon, which can be referred to as "citation orchestration" occurs when authors deliberately include citations not solely to endorse relevant research but rather to boost the reputation or visibility of specific researchers or their affiliations~\cite{didegah2013factors}.

The $h$-index, for example, can be manipulated by self-citations placed strategically to boost it~\cite{10.1007/s11192-010-0306-5}. Self-citations may be the most superficial and readily detectable type of manipulation and some authors reach extremely high levels of self-citations, \eg self-citations may account for more than $50\%$ or even $80\%$ of the citations that they receive~\cite{van_noorden,10.1371/journal.pbio.3000918}. However, orchestration may also involve more subtle citation cartels: a small number of scientists may orchestrate to cite each other, regardless of whether they are co-authors or not~\cite{10.3389/fphy.2016.00049,perez2019network}. Finally, with the advent of team science, it is becoming increasingly common for many teams to have a very large number of authors on each of the massive number of papers that they publish~\cite{papatheodorou2008inflated}. 
Such large-scale orchestration requires an in-depth analysis of the co-authorship patterns. 
All these behaviors and authorships and citation patterns are important to document so as to avoid a naive adoption of metrics like the $h$-index~\cite{bornmann2009state,ding2020exploring}.  

Eventually, citation orchestration can distort citation rankings and compromise the fairness and accuracy of research evaluation processes~\cite{waltman2011towards}.
In order to understand the extent of citation orchestration and characterize its presence across science, here we use the entire Scopus database, which provides comprehensive coverage of scholarly literature across diverse disciplines and facilitates a multi-dimensional exploration of citation patterns~\cite{burnham2006scopus,falagas2008comparison}.
We develop three readily accessible and interpretable metrics that may provide hints to possible citation orchestration behavior.
We also explore whether an excessive proportion of authors with extreme values in such orchestration indicators have had publications retracted from the literature for reasons other than publisher or journal error.

\section{Methods}\label{sec:methodology}

\subsection{Proposed Metrics}\label{subsec:metrics}
The metrics are defined at the authors' level.
We propose three different metrics that may offer hints for the presence of inflated $h$-indices due to small-scale or large-scale citation orchestration:

\paragraph{$\mathbf{C/h^2}$:} The first metric is equal to the number of citations divided by the square of the author's $h$-index.
A small value of $C/h^2$ indicates that the citations of the author are coming from papers that explicitly increase the $h$-index. 
An extremely small value for this metric may suggest that the citations of the author may have been strategically placed in order to increase the $h$-index.
This strategic placement may have been done either by the author himself/herself (self-citations) or by a small number of scientists who act as a coordinated citation cartel.
 
\paragraph{$\mathbf{A_{50\%C}}$:}
This metric is equal to the number of citing authors who cumulatively explain $50\%$ of the citations of an author.
An extremely small value of this metric may highlight either small-scale or large-scale orchestration.
Small-scale orchestration results in a low $A_{50\%C}$ when a single author either cites himself/herself or is cited by some specific people within a citation cartel in order to promote his or her $h$-index.
An extremely low $A_{50\%C}$ may be seen also with large-scale orchestration when a tightly-knit team effort dominates the scientific discipline and all/most authors are cited in all papers, a situation typical of nuclear and particle physics work affiliated with the European Organization of Nuclear Research.
To calculate $A_{50\%C}$, we take into account the citing author who contributes the most citations to the examined author. 
Once this citing author $A_1$ is identified, we re-sort the remaining citing authors based on the remaining unexamined citing papers (\ie excluding citing papers authored by $A_1$).
The highest contributor $A_2$ is identified then and similarly removed. The process is iterated until citing author $A_i$, when at least $50\%$ of the citations are explained.
This iterative process allows us to determine the number of citing authors who collectively explain at least $50\%$ of the examined authors' citations while avoiding double-counting of citations from the same papers. 

\paragraph{$\mathbf{A_{50}}$:}
The metric is equal to the number of co-authors with shared co-authorship exceeding $50$ papers. 
If this number is large, this may highlight large-scale orchestration, where the authors belong to a big cluster of authors that heavily cite one another in work that they do jointly. 
Typically, this reflects the situation where prolific teams publish papers with massive co-authorship, and then these papers massively cite papers produced by the same team.
This pattern is overwhelmingly seen in work done by the European Organization of Nuclear Research but this type of practice may also be appearing increasingly also in other scientific fields~\cite{cronin2001hyperauthorship}.

\subsection{Scopus Dataset}
As part of the project, we were given access to the International Center for the Study of Research (ICSR) Lab\footnote{\url{https://www.elsevier.com/icsr/icsrlab}}, a cloud-based computational platform that allows its users to analyze large structured datasets such as the publication metadata of Scopus\footnote{\url{https://www.elsevier.com/solutions/scopus/how-scopus-works/metrics}}.
The platform provides access to the publication metadata of $95,242,081$ papers in total as of $12^{th}$ of March 2023.
These papers are authored by $49,255,352$ different author IDs in total.
The precision and accuracy of the author ID files in Scopus have been well characterized~\cite{kawashima2015accuracy}; while some authors have their papers split into more than one ID file and some specific ID files contain papers by more than one author, the errors are very small to materially affect the distributions of the proposed metrics, and may be even less influential for the extreme tails that are likely to be most informative for orchestration.

\subsection{Preprocessing}
We only consider publications that are full papers, \ie fall into the Scopus categories of articles, conference papers, or reviews, in concordance with previous work~\cite{ioannidis2018thousands}, with a total of $82,694,786$ eligible papers. We next determine the field and subfield of study for each author using the Science-Metrix\footnote{\url{https://science-metrix.com/}} \cite{archambault2011towards} classification. 
Each publication is assigned to one of 22 fields and one of 174 subfields. 
In case an author has published multiple papers and those papers belong to different fields, we choose the one where most of the author's papers are assigned. 
In case of ties, we choose the field whose papers have received the largest number of citations. 
If there are still ties, we randomly select one of the fields. $2,889,779$ authors without specified subfield or field are excluded from all analyses. 
Table~\ref{tab:field_authors} shows the number of authors eligible for analysis in each field.

\begin{table*}[t]
    \begin{center}
    \scriptsize
    \renewcommand{\arraystretch}{0.9}
    \begin{tabular}{|l|c|c|c|c|} 
        \hline Field of study & Number of authors & Percentage & Number of authors analyzed & Percentage\\
        \hline
        Agriculture, Fisheries \& Forestry & $1,446,173$ & $3.12 \%$ & $35,839$ & $2.39 \%$\\
        Biology & $1,586,467$ & $3.42 \%$ & $64,737$ & $4.33\%$\\
        Biomedical Research & $3,057,457$ & $6.60 \%$ & $189,457$ & $12.66\%$\\
        Built Environment \& Design & $273,254$ & $0.60 \%$ & $3,638$ & $0.24\%$\\
        Chemistry & $2,464,113$ & $5.31 \%$ & $91,721$ & $6.13\%$\\
        Clinical Medicine & $15,594,844$ & $33.63 \%$ & $579,366$ & $38.71\%$\\
        Communication \& Textual Studies & $328,254$ & $0.71 \%$ & $1,617$ & $0.11\%$\\  
        Earth \& Environmental Sciences & $1,345,726$ & $2.90 \%$ & $58,067$ & $3.88\%$\\
        Economics \& Business & $909,251$ & $2.00\%$ & $20,776$ & $1.39\%$\\
        Enabling \& Strategic Technologies & $5,103,262$ & $11.00 \%$ & $101,820$ & $6.80\%$\\ 
        Engineering & $3,679,238$ & $7.94\%$ & $56,200$ & $3.75\%$\\ 
        General Arts, Humanities \& Social Sciences & $5,992$ & 
        $0.01 \%$ & $2$ & 
        $0.00 \%$\\ 
        General Science \& Technology & $30,121$ & $0.06 \%$ & $77$ & $0.01 \%$\\
        Historical Studies & $241,666$ & $0.52\%$ & $1,550$ & $0.10\%$\\
        Information \& Communication Technologies & $4,094,210$ & $8.83\%$ & $57,163$ & $3.82\%$\\
        Mathematics \& Statistics & $410,548$ & $0.89\%$ & $9,605$ & $0.64\%$\\
        Philosophy \& Theology & $111,270$ & $0.24\%$ & $433$ & $0.03\%$\\
        Physics \& Astronomy & $2,649,145$ & $5.71\%$ & $174,028$ & $11.63\%$\\
        Psychology \& Cognitive Sciences & $487,914$ & $1.10\%$ & $18,264$ & $1.22\%$\\
        Public Health \& Health Services & $1,221,920$ & $2.64\%$ & $19,928$ & $1.33\%$\\
        Social Sciences & $1,300,068$ & $2.80\%$ & $12,372$ & $0.83\%$\\
        Visual \& Performing Arts & $25,800$ & $0.06\%$ & $20$ & $0.00\%$\\        
        \hline
    \end{tabular}
    \end{center}
    \caption{Number and percentage of total Scopus author IDs and of Scopus author IDs eligible for analysis (those with at least 5 full papers and at least 1000 citations).}
    \label{tab:field_authors}
\end{table*}

We focus on a subset of authors who have made a significant contribution to the research community and satisfy the following two conditions: (1) they have published more than $5$ full papers, and (2) their papers have received at least $1,000$ citations in total. 
Eventually, $1,496,680$ authors satisfy the above conditions. 
Their numbers per scientific field are also shown in Table \ref{tab:field_authors}.

\subsection{Analyses of distributions for proposed indicators}
For each of the three proposed indicators, we generate the distribution of values for the $1,496,680$ eligible authors. We focus on the left-side tails (very small values) for the first two indicators and the right-side tails for the third one. There is no absolute cut-off that can separate authors where orchestration has been done on purpose or may be considered to be substantial. The most extreme values are more likely to reflect gaming, but gaming may have happened to a various extend also for citation profiles of authors that have less extreme values. Here we focus primarily on the lowest $1\%$ percentile for $C/h^2$ and $A_{50\%C}$ and the highest $1\%$ for $A_{50}$ metric. We also present values for the $5\%$ threshold as well as median and interquartile range values for each of the three indicators. We also show how extreme $1\%$ percentile authors are distributed across the main fields of
science, so as to demonstrate whether these patterns are enriched in specific fields. For $A_{50\%C}$ and $A_{50}$ specifically, we provide analyses that exclude the authors categorized in the field of Physics and Astronomy, since their tails of interest are otherwise overwhelmed by nuclear and particle physics authors in work affiliated with the European Organization for Nuclear Research.

\subsection{Coexistence of orchestration indicators}
Furthermore, we explore whether the three indicators tend to co-exist or tend to be mutually exclusive from each other by examining their pair-wise co-existence with odds ratios and $95\%$ confidence intervals and p-values based on chi-square test.  We focus on $1,322,652$ authors remaining after excluding those in the field of Physics \& Astronomy who are known to often have massive co-authorship patterns.

\subsection{Highly cited status among authors with orchestration indicators}
We analyze how many authors with orchestration indicators are listed in the science-wide author database of standardized citation indicators \footnote{\url{https://elsevier.digitalcommonsdata.com/datasets/btchxktzyw/6}}. This database includes highly cited authors, selected based on either being among the top 100,000 scientists across all science or being among the top-2\% of the scientists in the same subfield according to a composite score that considers 6 citation indicators (with and without self-citations). The publicly available database includes 204,643 authors meeting these criteria in a snapshot from Scopus as of October 2023, that reflects citations up to the end of 2022. To match our author set with the standardized citation database, we used exact string matching based on full name (concatenation of last name and all initials, normalized to lowercase and punctuation-stripped). When multiple matches occurred, we retained only exact matches for name; otherwise, we excluded ambiguous cases. This conservative approach minimizes false positives
and ensures reproducibility, although it may result in some false negatives for authors that are not matched because of minor irregularities in the two datasets. The composite score prioritizes impact (citations) over productivity (number of publications) and takes into account factors such as co-authorship and author positions (\eg, single, first, or last author)~\cite{10.1371/journal.pbio.1002501, Ioannidis2019, 10.1371/journal.pbio.3000918}.

\subsection{Analysis of retractions}
Finally, we explore whether authors with extreme values of orchestration metrics have had at least one retracted publication. For this analysis, we exclude again Physics \& Astronomy authors. We select a random sample of 100 authors from each group of authors defined by the presence of an extreme value for only one of the three indicators or for two of the three indicators. 5 of the 6 groups had more than 100 authors, and thus 100 were selected using simple random sampling without replacement, ensuring that each author in the group had an equal probability of being selected. The group of those with extremely low values in $C/h^2$ and extremely high values of $A_{50}$ included only 11 authors, so all of them are assessed. Searches in the Retraction Watch database \footnote{\url{https://retractiondatabase.org/}} were performed in June 2025. We did not consider expressions of concern or corrections in the absence of full retraction. We did not consider also retractions with replacement/republication. Finally, we did not consider retractions that were clearly due to an error of the publisher or journal, with no error on the part of the authors.

\section{Results}

\subsection{Distributions of indicators}
\begin{figure*}
     \centering
     \begin{subfigure}[b]{0.45\linewidth}
         \centering
         \includegraphics[width=\linewidth]{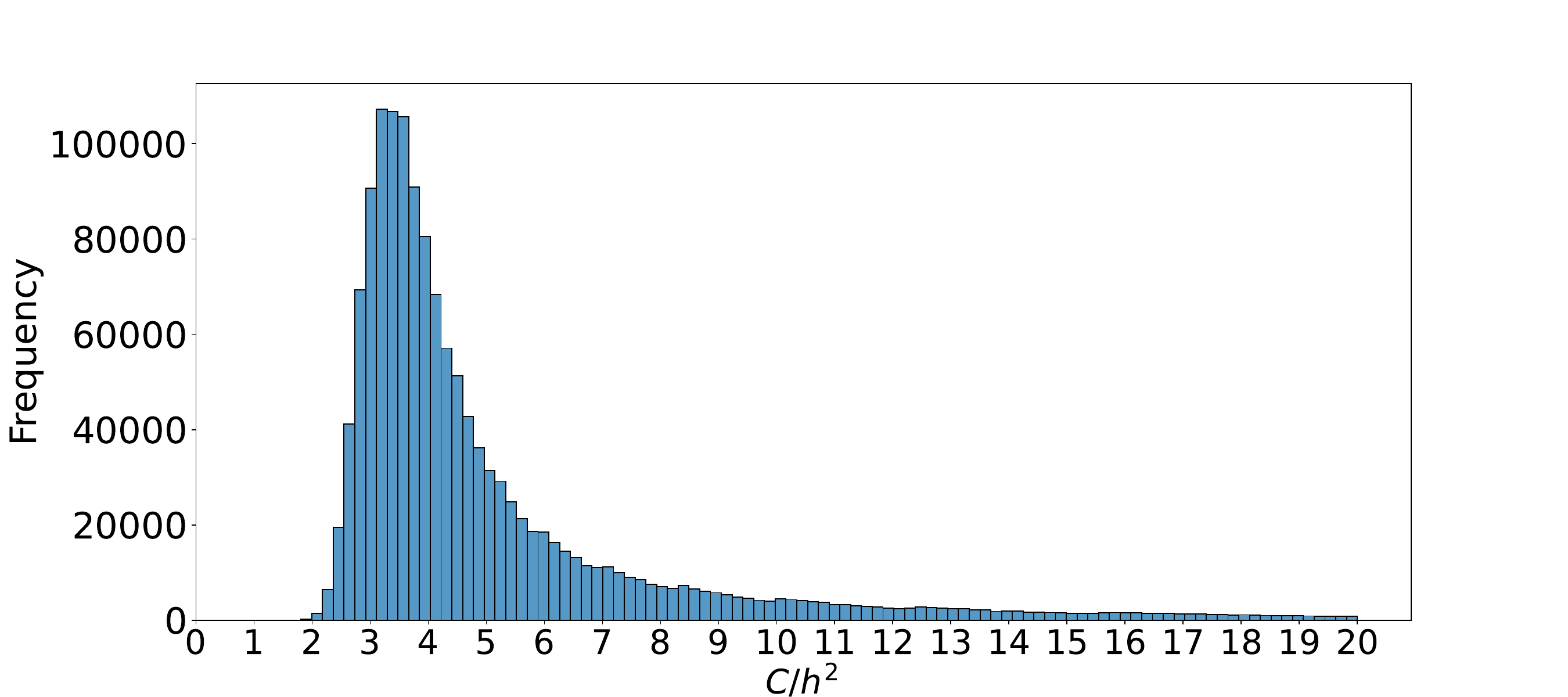}
         \caption{The distribution of the metric $C/h^2$ (authors with values above 20 are not shown)}
         \label{fig:elsevier_distr_a1}
     \end{subfigure}
     \hfill
     \begin{subfigure}[b]{0.45\linewidth}
         \centering
         \includegraphics[width=\linewidth]{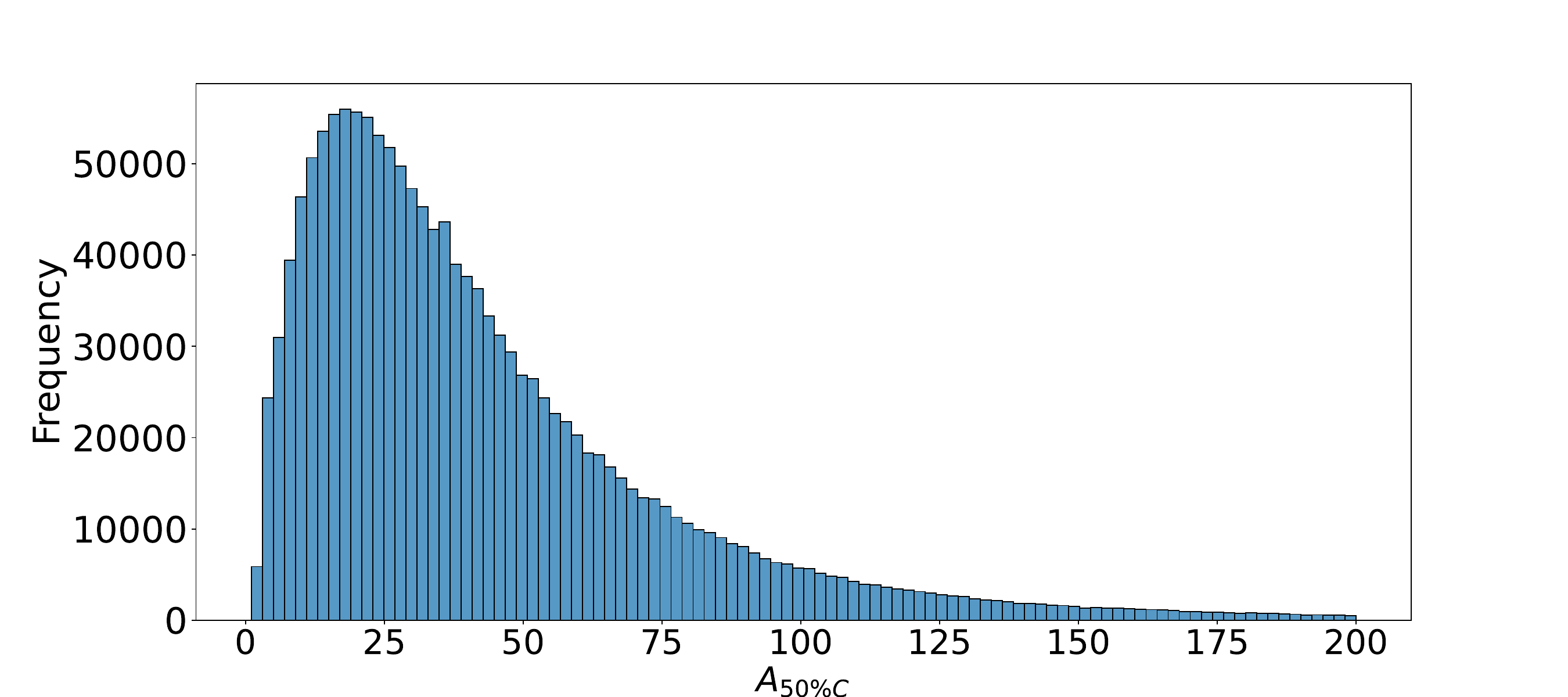}
         \caption{The distribution of the metric $A_{50\%C}$ (authors with values above 200 are not shown)}
         \label{fig:elsevier_distr_a7_all}
     \end{subfigure}
     \hfill
     \begin{subfigure}[b]{0.45\linewidth}
         \centering
         \includegraphics[width=\linewidth]{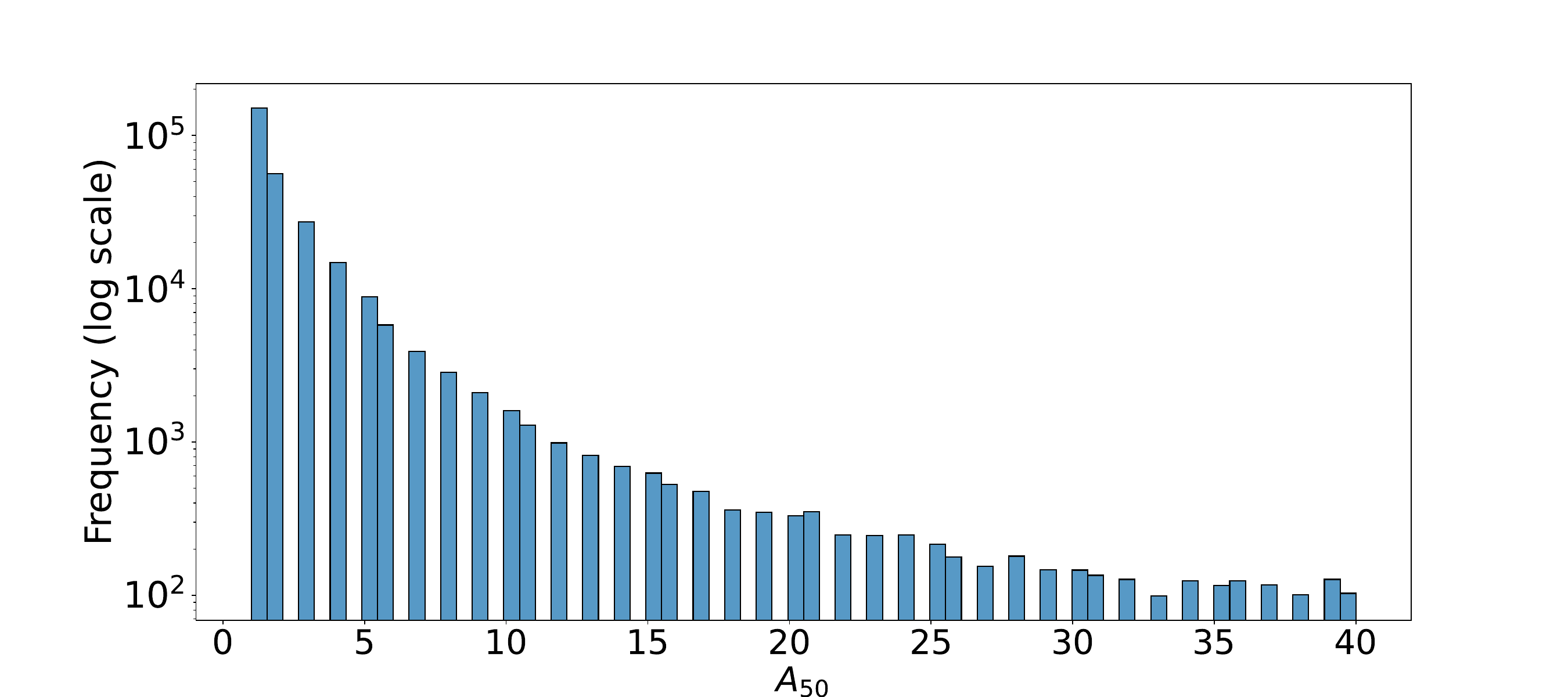}
         \caption{The distribution of the metric $A_{50}$ (authors with values of 0 and above 40 are not shown).}
         \label{fig:elsevier_distr_a6_all}
     \end{subfigure}
     \hfill
     \begin{subfigure}[b]{0.45\linewidth}
         \centering
         \includegraphics[width=\linewidth]{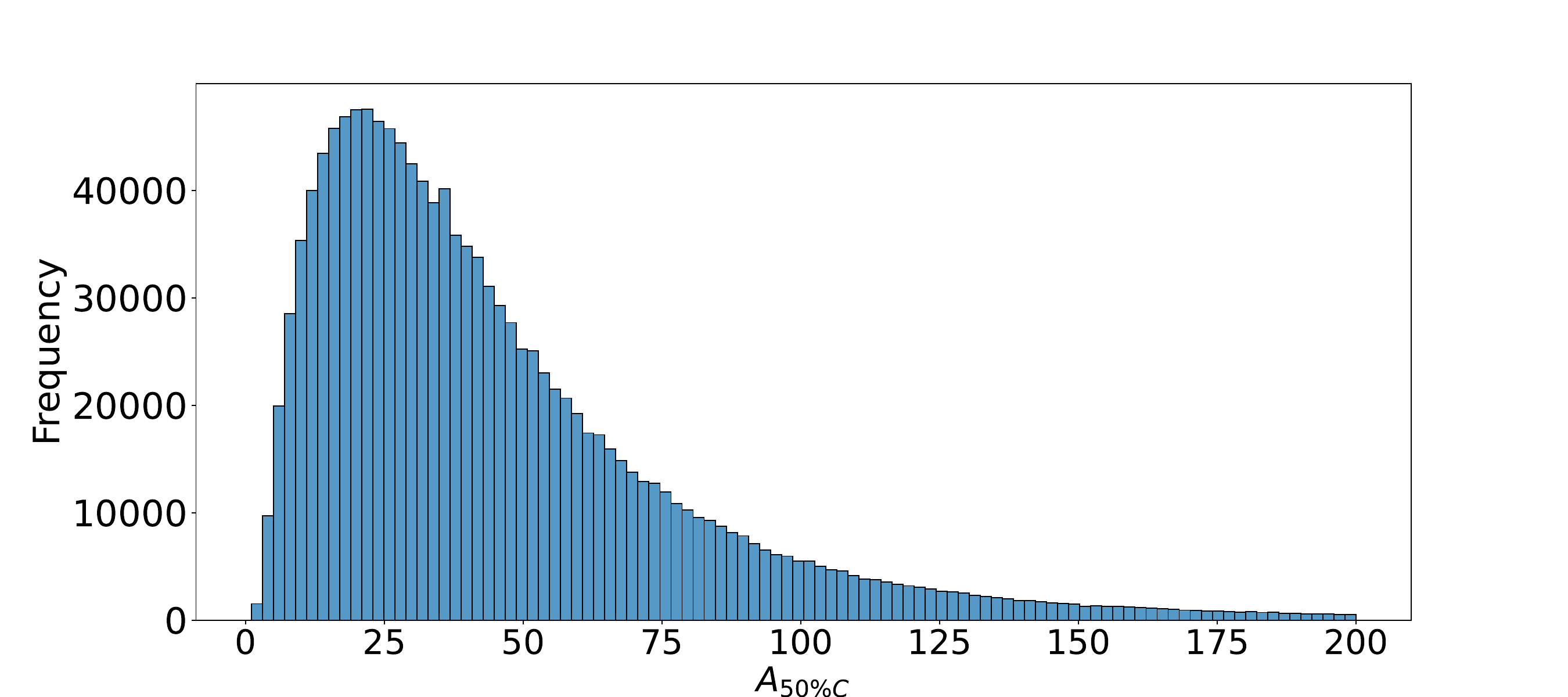}
         \caption{The distribution of the metric $A_{50\%C}$ (excluded authors with values exceeding 200 and those in the field of Physics and Astronomy).}
         \label{fig:elsevier_distr_a7}
     \end{subfigure}
     \hfill
     \begin{subfigure}[b]{0.45\linewidth}
         \centering
         \includegraphics[width=\linewidth]{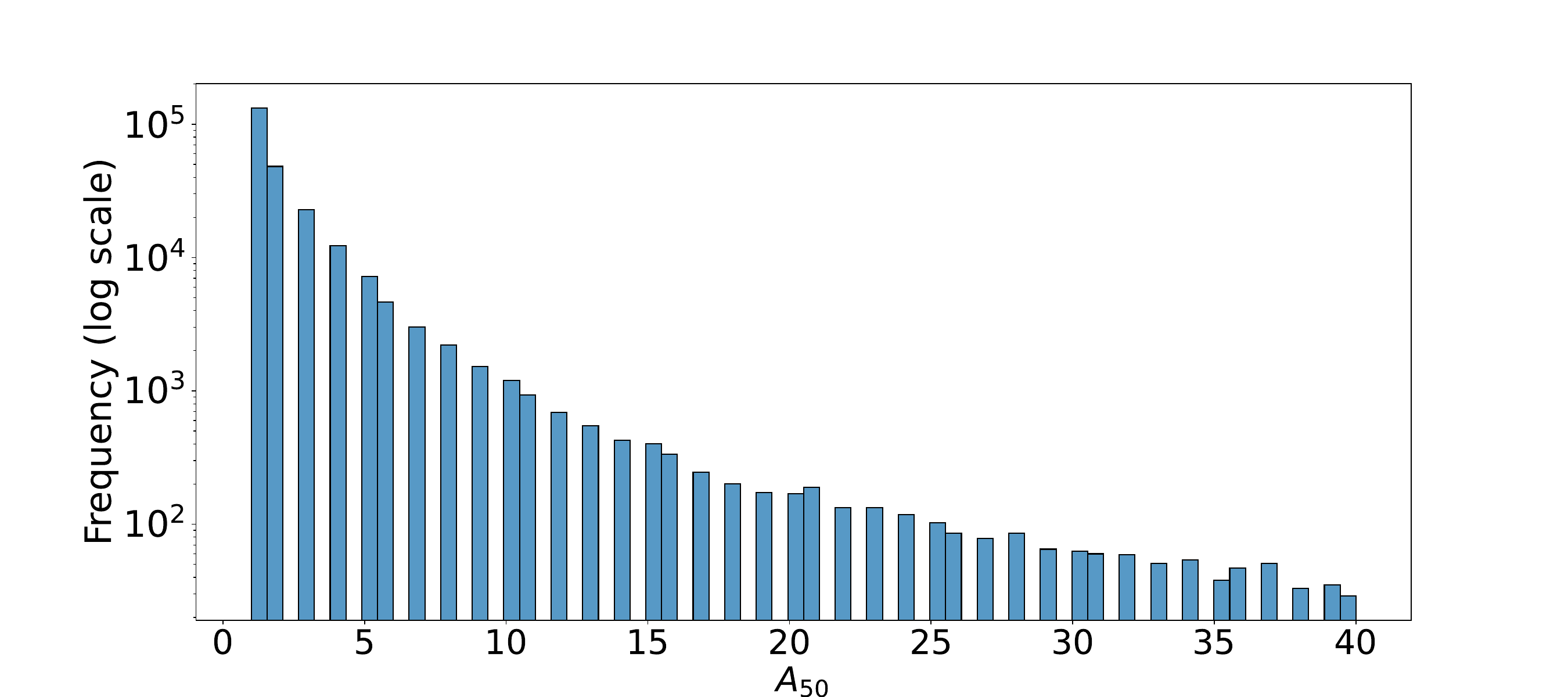}
         \caption{The distribution of the metric $A_{50}$ (excluded authors with values above 40 and those in the field of Physics and Astronomy).}
         \label{fig:elsevier_distr_a6}
     \end{subfigure}
        \caption{The distribution of all the proposed metrics}
        \label{fig:elsevier_distr}
\end{figure*}

\begin{figure*}
     \centering
     \begin{subfigure}[b]{0.45\linewidth}
         \centering
         \includegraphics[width=\linewidth]{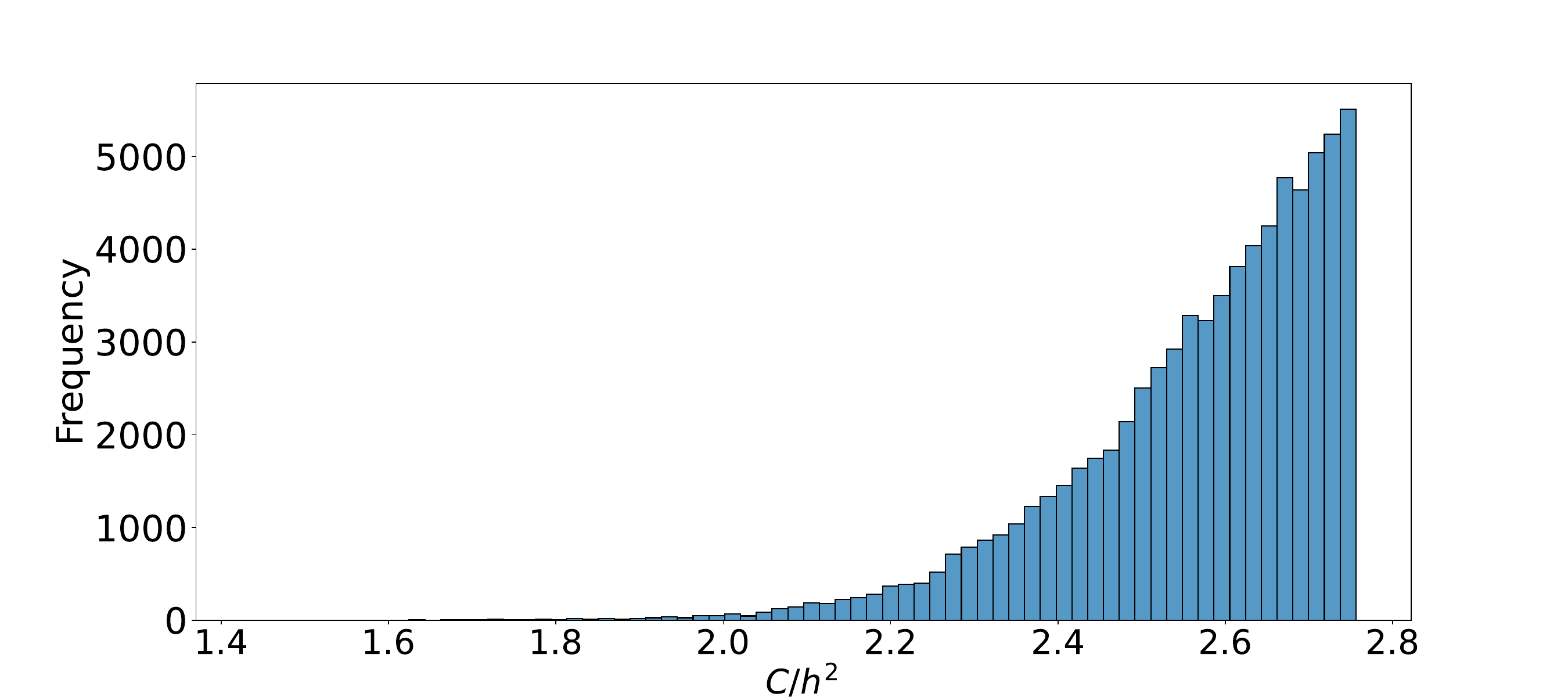}
         \caption{The distribution of the metric $C/h^2$ on the lowest 5\%}
         \label{fig:elsevier_distr_a1_zoom}
     \end{subfigure}
     \hfill
     \begin{subfigure}[b]{0.45\linewidth}
         \centering
         \includegraphics[width=\linewidth]{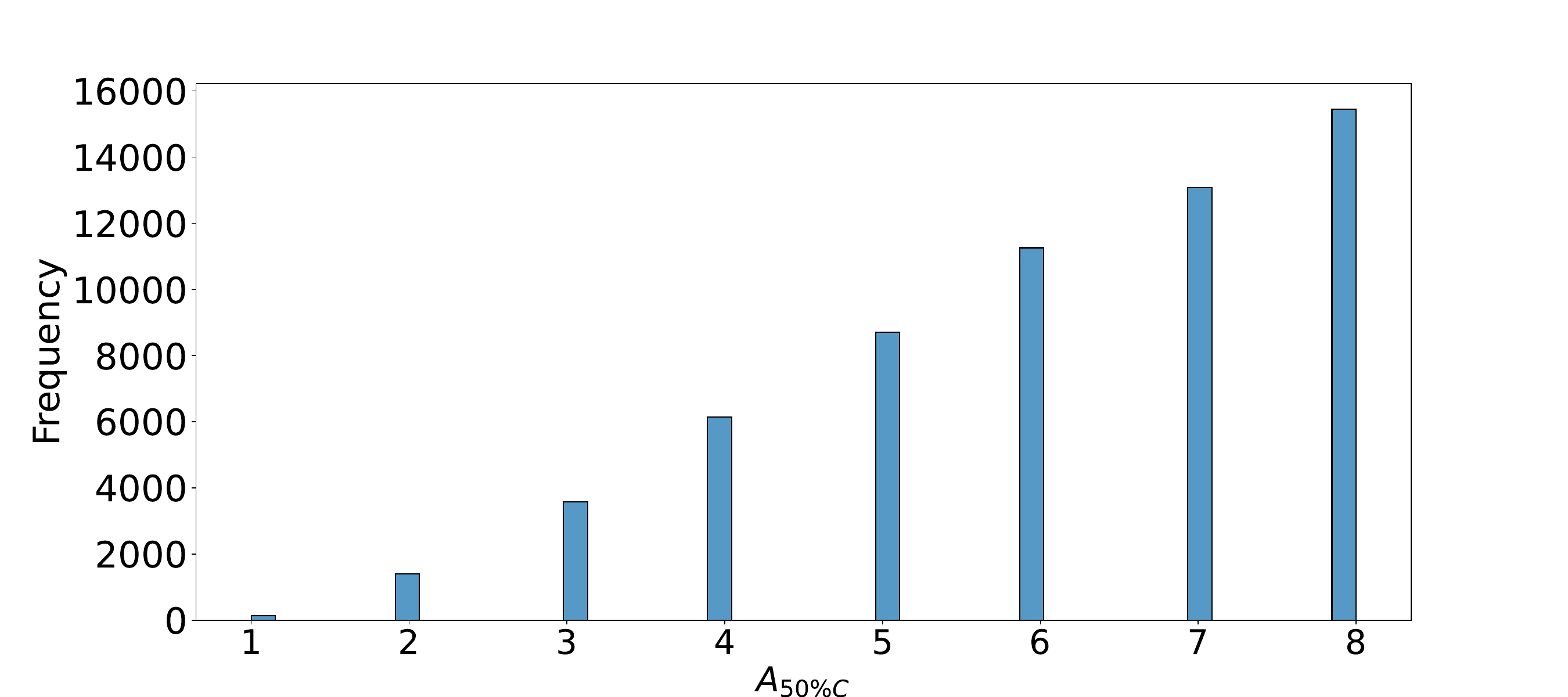}
         \caption{The distribution of the metric $A_{50\%C}$ on the lowest 5\%}
         \label{fig:elsevier_distr_a7_zoom}
     \end{subfigure}
     \hfill
     \begin{subfigure}[b]{0.45\linewidth}
         \centering
         \includegraphics[width=\linewidth]{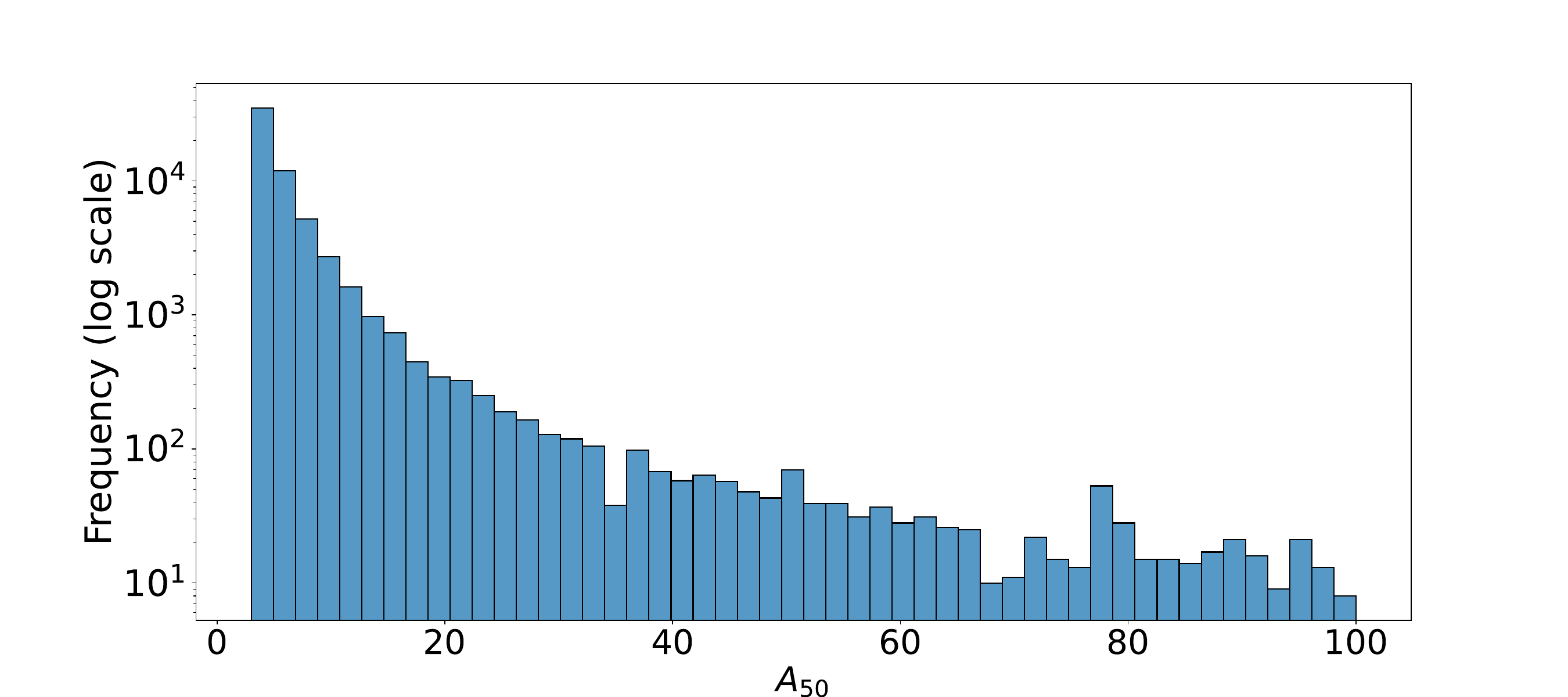}
         \caption{The distribution of the metric $A_{50}$ on the upper 5\% (authors with values above 100 are not shown)}
         \label{fig:elsevier_distr_a6_zoom}
     \end{subfigure}
        \caption{The distribution of all the proposed metrics on the lowest or upper 5\%}
        \label{fig:elsevier_distr_zoom}
\end{figure*}

The distribution of the three proposed indicators appears in Figure~\ref{fig:elsevier_distr} and Figure~\ref{fig:elsevier_distr_zoom} zooms out specifically on the lower $5\%$ percentile of the first two indicators and the upper $5\%$ percentile of the third orchestration indicator. The distributions are right-skewed.
For the $A_{50\%C}$ and $A_{50}$ metrics, we provide also distributions excluding the field of Physics and Astronomy so as to allow revealing better the patterns for all other fields.

\subsection{Metric \texorpdfstring{$C/h^2$}{C/h2}}
For the metric $C/h^2$, the $1^{st}$ percentile value is $2.45$ and the $5^{th}$ percentile value is $2.76$. Comparatively, the median is $4.11$ and the interquartile range is $3.36$ to $6.11$. Here, we focus on the 1\% percentile ($14,967$ authors) that may have the strongest hint for small-scale orchestration. Table~\ref{tab:field_authors_perc} shows that the scientific field allocation of these $14,967$ authors is largely similar to the allocation of all examined authors, with some exceptions. Specifically, there are proportionally more
than $1.5$-fold higher representation of authors with extremely low $C/h^2$ values compared to all examined authors in 3 fields: Agriculture, Fisheries and Forestry, Chemistry, and Earth and Environmental Sciences.

\begin{table*}[t]
\small
\centering
\resizebox{\textwidth}{!}{
    \begin{tabular}{l|p{2cm}|p{3cm}|p{3cm}|p{3cm}} 
        \hline Field of study & Percentage of examined authors & Percentage of lowest 1\% authors of $C/h^2$ metric & Percentage of lowest 1\% authors of metric $A_{50\%C}$ (excluding Physics \& Astronomy) & Percentage of upper 1\% authors of metric $A_{50}$ (excluding Physics \& Astronomy)\\
        \hline
        Agriculture, Fisheries \& Forestry & $2.39\%$ & $5.74\%$ & $2.19\%$ & $1.43\%$\\
        Biology & $4.33\%$ & $5.99\%$ & $4.34\%$ & $0.75\%$\\
        Biomedical Research & $12.66\%$ & $12.19\%$ & $4.76\%$ & $9.03\%$ \\
        Built Environment \& Design & $0.24\%$ & $0.25\%$ & $0.44\%$ & $0.02\%$\\
        Chemistry & $6.13\%$ & $13.80\%$ & $15.47\%$ & $3.07\%$\\
        Clinical Medicine & $38.71\%$ & $29.20\%$ & $17.09\%$ & $73.29\%$\\
        Communication \& Textual Studies & $0.11\%$ & $0.03\%$ & $0.26\%$ & $0.00\%$\\  
        Earth \& Environmental Sciences & $3.88\%$ & $5.93\%$ & $4.82\%$ & $1.34\%$\\
        Economics \& Business & $1.39\%$ & $0.62\%$ & $2.13\%$ & $0.00\%$\\
        Enabling \& Strategic Technologies & $6.80\%$ & $7.96\%$ & $10.84\%$ & $7.31\%$\\ 
        Engineering & $3.75\%$  & $4.64\%$ & $12.18\%$ & $1.17\%$ \\ 
        General Arts, Humanities \& Social Sciences & 
        $0.00 \%$ & $0.00 \%$ & $0.00 \%$ & $0.00\%$ \\ 
        General Science \& Technology & $0.01 \%$ & $0.00 \%$ & $0.00 \%$ & $0.00 \%$\\
        Historical Studies & $0.10\%$ & $0.08\%$ & $0.24\%$ & $0.00\%$\\
        Information \& Communication Technologies & $3.82\%$ & $1.06\%$ & $11.00\%$ & $1.59\%$\\
        Mathematics \& Statistics & $0.64\%$ & $0.48\%$ & $9.44\%$ & $0.04\%$\\
        Philosophy \& Theology & $0.03\%$ & $0.00\%$ & $0.16\%$ & $0.00\%$ \\
        Physics \& Astronomy & $11.63\%$ & $9.77\%$ & $-$ & $-$\\
        Psychology \& Cognitive Sciences & $1.22\%$ & $0.84\%$ & $1.99\%$ & $0.25\%$ \\
        Public Health \& Health Services & $1.33\%$ & $0.92\%$ & $0.96\%$ & $0.67\%$\\
        Social Sciences & $0.83\%$ & $0.46\%$ & $1.68\%$ & $0.03\%$ \\
        Visual \& Performing Arts & $0.00\%$ & $0.00\%$ & $0.02\%$ & $0.00\%$\\        
        \hline
    \end{tabular}}
    \caption{Authors with extreme metrics: Distribution across the 22 main fields of science.}
    \label{tab:field_authors_perc}
\end{table*}


\begin{figure*}
     \centering
     \begin{subfigure}[b]{0.45\textwidth}
         \centering
         \includegraphics[width=\textwidth]{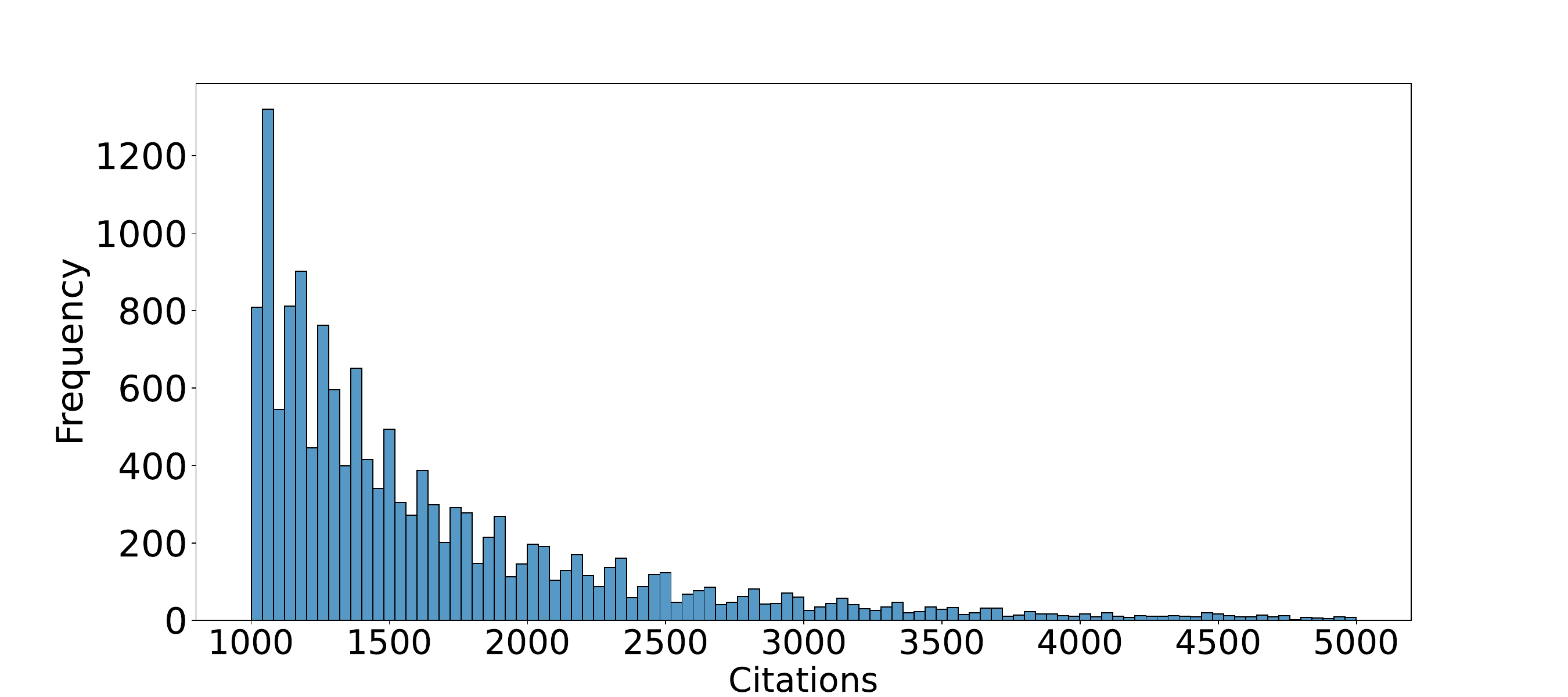}
        \caption{Distribution of the citations, considering only whoever author has an $C/h^2$ value less than $2.45$ (authors with citations above $5,000$ are not shown).}
        \label{fig:elsevier_distr_cit_a1}
     \end{subfigure}
     \hfill
     \begin{subfigure}[b]{0.45\textwidth}
         \centering
         \includegraphics[width=\textwidth]{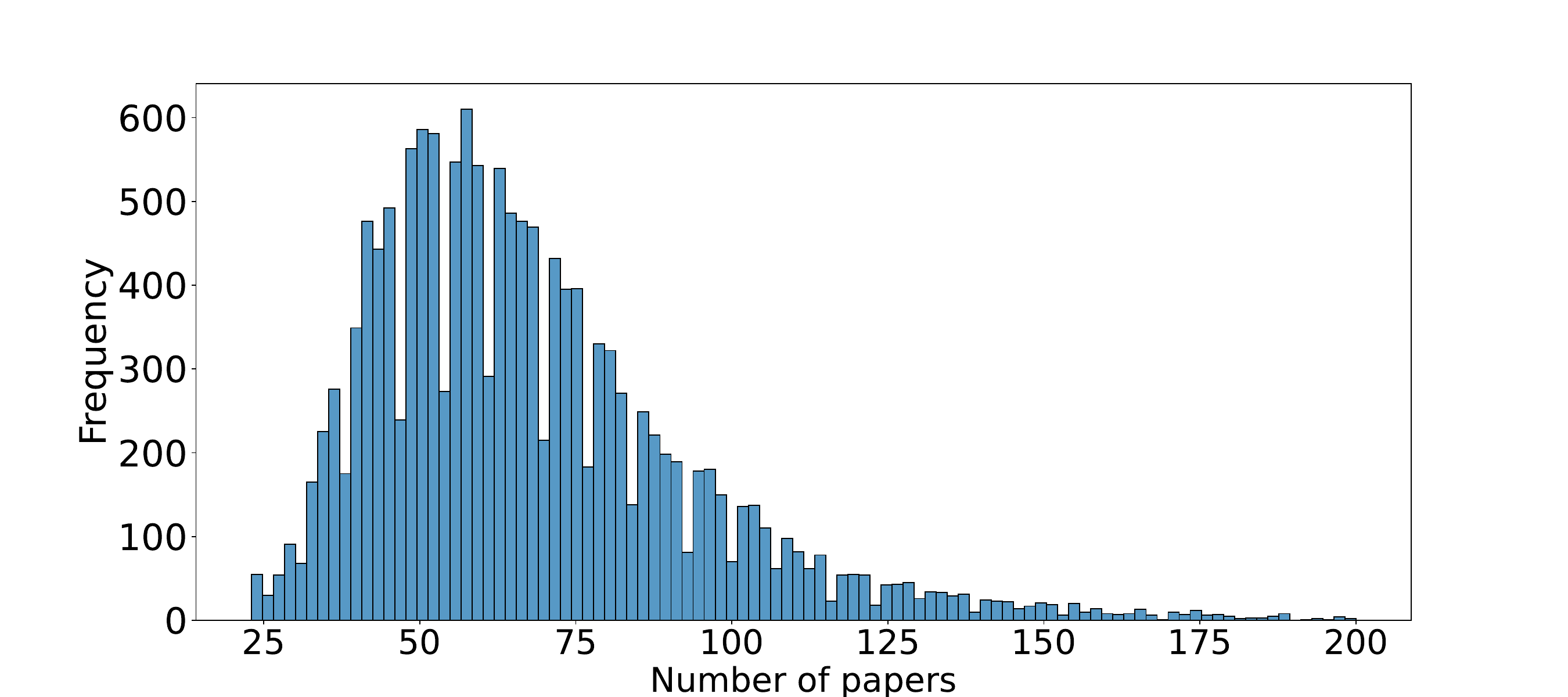}
        \caption{The distribution of the number of papers published by authors with an $C/h^2$ value on the lowest 1\% (authors with more than 200 papers are not shown).}
        \label{fig:elsevier_distr_numPapers_a1}
     \end{subfigure}
     \hfill
     \begin{subfigure}[b]{0.45\textwidth}
         \centering
         \includegraphics[width=\textwidth]{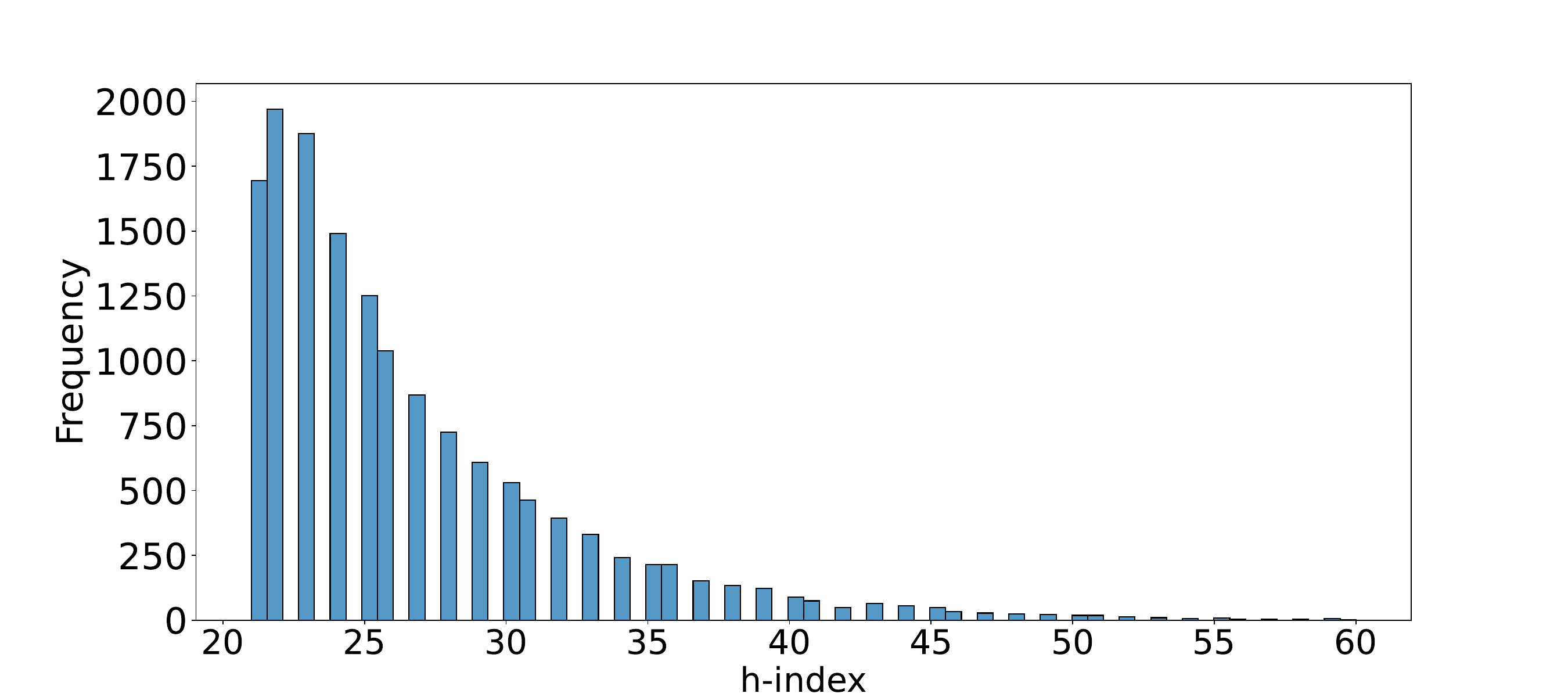}
        \caption{Distribution of the $h$-index of the authors that have an $C/h^2$ value less than $2.45$ (authors with $h$-index above 60 are not shown).
        }
        \label{fig:elsevier_distr_hindex_a1}
     \end{subfigure}
        \caption{The distribution of citations, number of papers and $h$-index of the authors that belong to the lowest 1\% of metric $C/h^2$.}\label{fig:elsevier_cit_hindex_papers_a1}
\end{figure*}

Figure~\ref{fig:elsevier_cit_hindex_papers_a1} depicts the distribution of citations, the number of published papers, and the $h$-index, respectively, for this group of $14,717$ authors with the lowest $C/h^2$ metric. These are productive authors (median $63$ papers, the interquartile range is $50$ to $81$ publications) and they all have $h$-indices exceeding $21$, while some of them have even very large $h$-indices.

\subsection{Metric \texorpdfstring{$A_{50\%C}$}{A50\%C}}

For the $A_{50\%C}$ metric, we observed that $73\%$ ($4,339$ out of $5,881$) of authors with the lowest $1\%$ percentile values of $A_{50\%C}$ belong to Physics \& Astronomy (of those, $3,808$ belong to the subfield of nuclear and particle physics). Therefore, we excluded Physics \& Astronomy authors in order to examine better the distribution for the rest of science. All results that follow retain analyses excluding Physics \& Astronomy.
For the $A_{50\%C}$ metric, the $1^{st}$ percentile is $5$ while the median is $36$, and the interquartile range is $21$ to $60$.
The $11,277$ authors who belong to this lowest $1\%$ percentile (metric value less than $5$) are more likely to engage in self-citation or be part of networks that highly cite each other.
Table~\ref{tab:field_authors_perc} shows the disciplines to which these $11,277$ authors belong.
There is a more than 1.5-fold higher representation of authors with exceptionally low $A_{50\%C}$ values compared to all examined authors in $10$ specific fields: Chemistry, Enabling \& Strategic Technologies,  Engineering, Historical Studies, Information \& Communication Technologies, Mathematics \& Statistics, Philosophy and Theology, Psychology and Cognitive Sciences, Social Sciences, and Visual and Performing Arts.

\begin{figure*}
     \centering
     \begin{subfigure}[b]{0.45\textwidth}
         \centering
         \includegraphics[width=\textwidth]{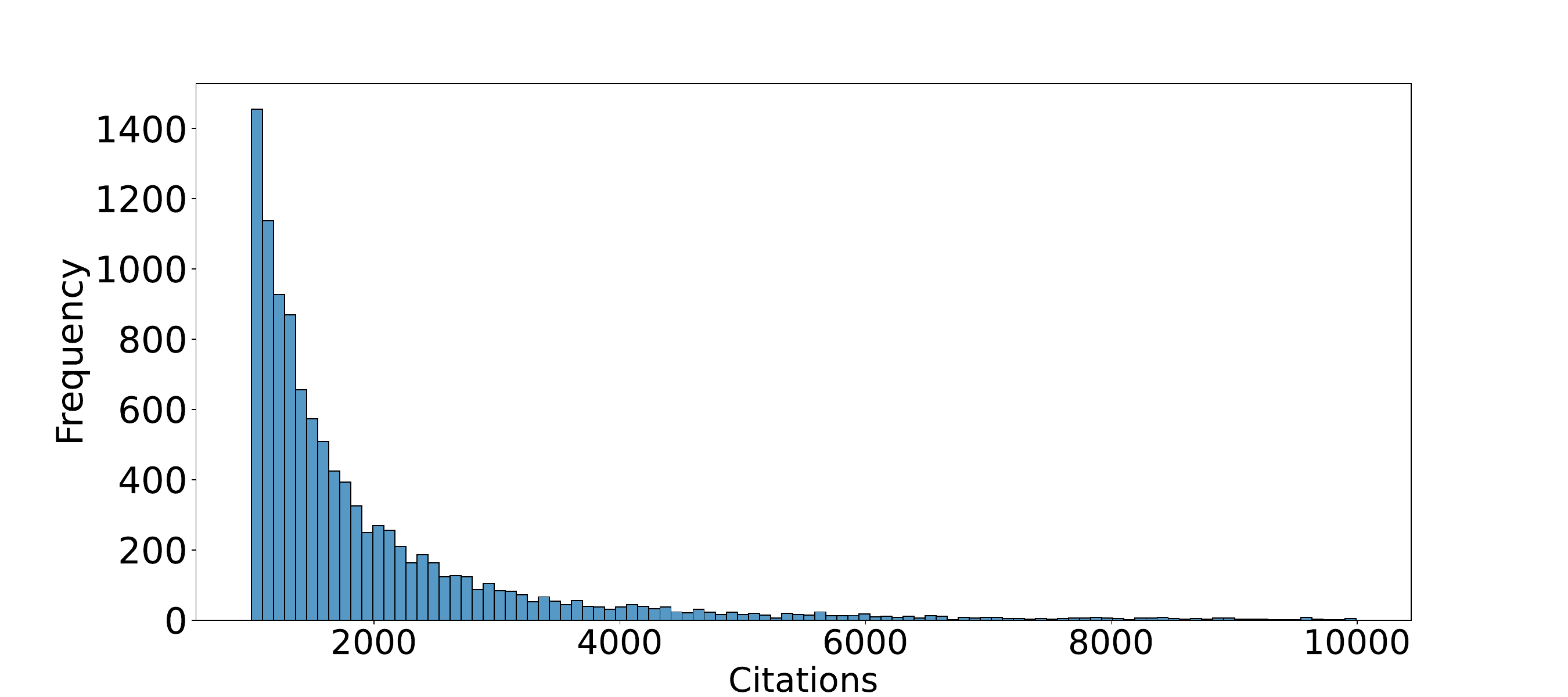}
        \caption{Distribution of the citations, considering only whoever author has an $A_{50\%C}$ value less than $5$ (authors with citations above $10,000$ are not shown).}
        \label{fig:elsevier_distr_cit_a7}
     \end{subfigure}
     \hfill
     \begin{subfigure}[b]{0.45\textwidth}
         \centering
         \includegraphics[width=\textwidth]{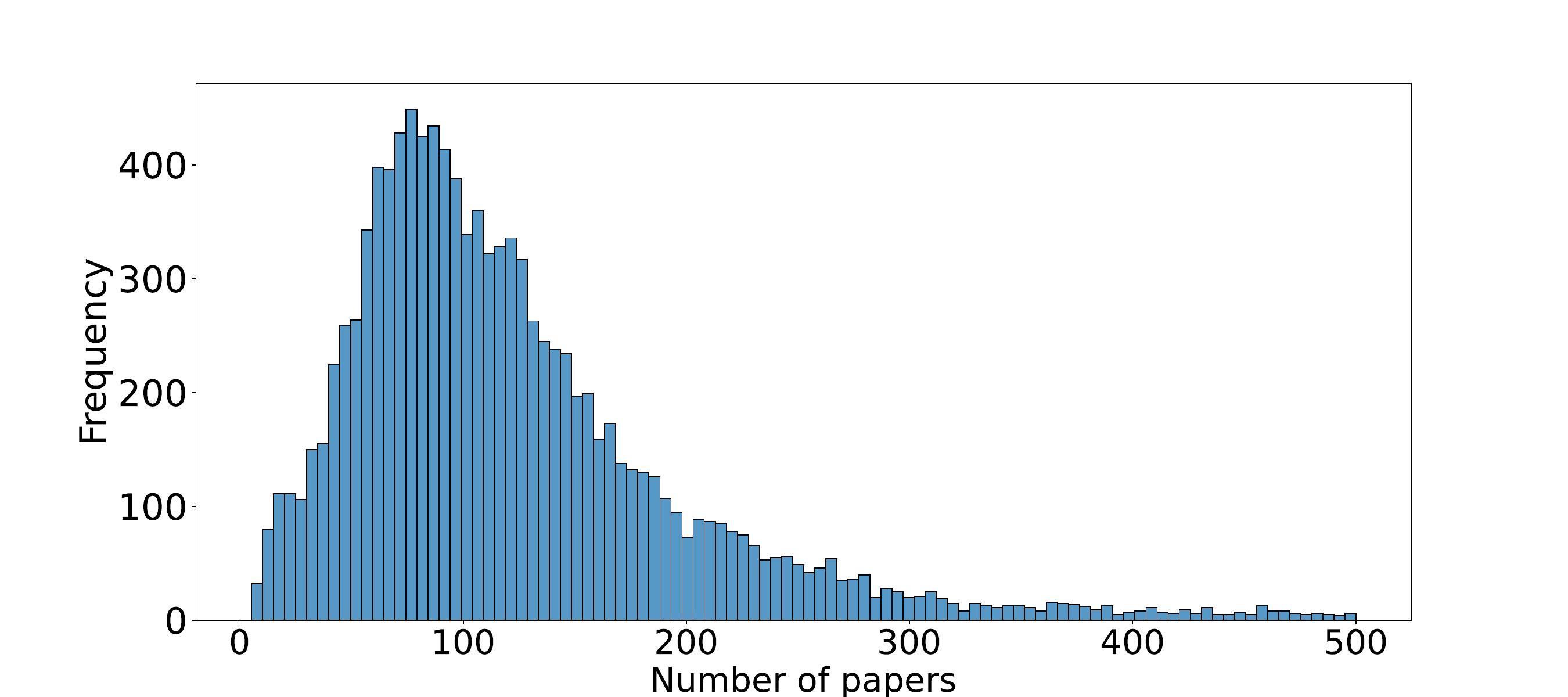}
        \caption{The distribution of the number of papers published by authors with an $A_{50\%C}$ value on the lowest 1\% (authors with more than $500$ papers are not shown).}
        \label{fig:elsevier_distr_numPapers_a7}
     \end{subfigure}
     \hfill
     \begin{subfigure}[b]{0.45\textwidth}
         \centering
         \includegraphics[width=\textwidth]{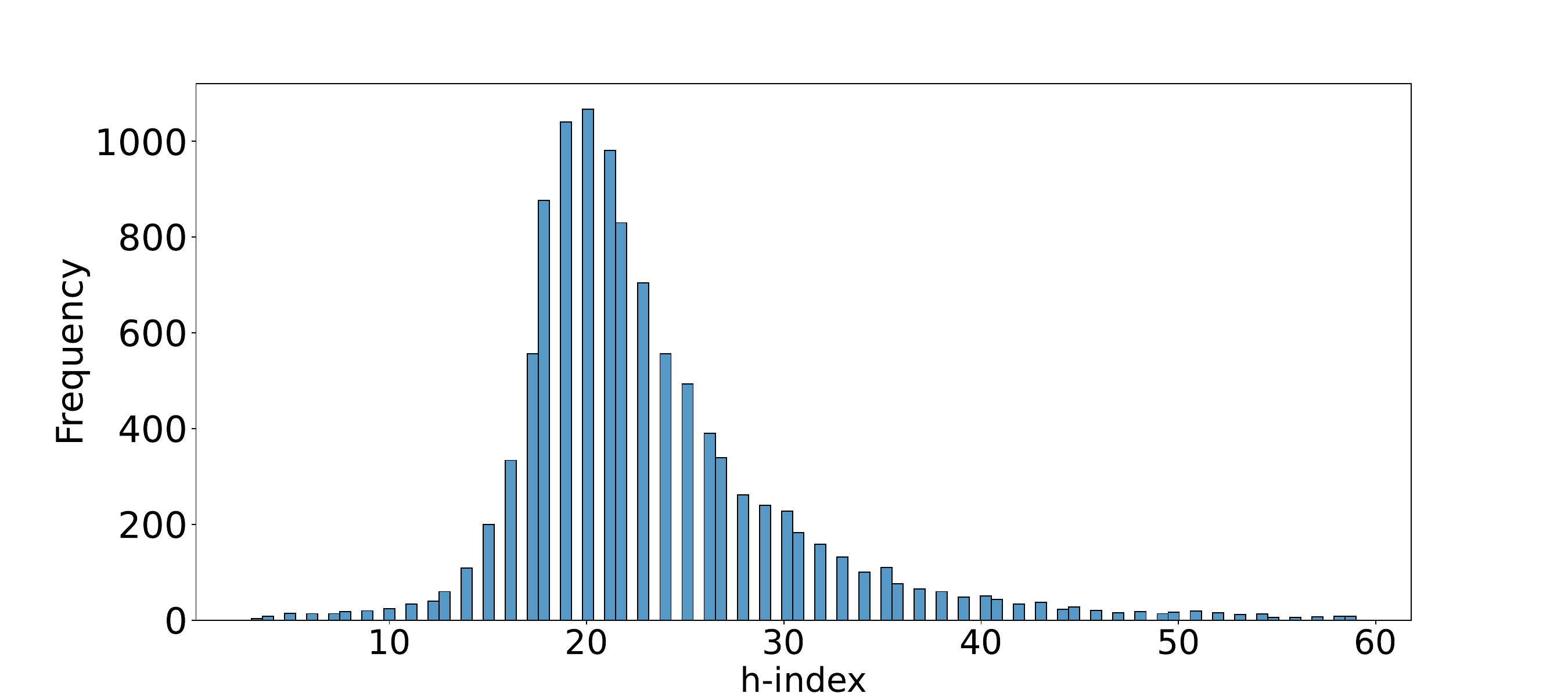}
        \caption{Distribution of the $h$-index of the authors that have an $A_{50\%C}$ value less than $5$ (authors with $h$-index above $60$ are not shown).
        }
        \label{fig:elsevier_distr_hindex_a7}
     \end{subfigure}
    \caption{The distribution of citations, number of papers and $h$-index of the authors that belong to the lowest 1\% of metric $A_{50\%C}$.}
    \label{fig:elsevier_cit_hindex_papers_a7}
\end{figure*}

Figure~\ref{fig:elsevier_cit_hindex_papers_a7} depicts the distribution of citations, the number of published papers, and the h-index, respectively, for this group of $11,277$ authors with the lowest $1\%$ percentile for the $A_{50\%C}$ metric. These are mostly productive authors (median
$104$ papers, interquartile range $71$ to $153$ publications) and they mostly ($59\%$ of these authors) have $h$-indices exceeding $20$, while some of them have even large $h$-indices.



\subsection{Metric \texorpdfstring{$A_{50}$}{A50}}\label{subsec:a6}

Metric $A_{50}$ aims to identify authors that heavily collaborate with each other, essentially functioning as a metric for detecting large-scale orchestration. 
Among the $1,496,680$ considered authors, we exclude the authors that belong to the field of "Physics \& Astronomy", as it is widely recognized for its consistent involvement in large-scale collaborative endeavors, leading to a potential bias in the analysis.
In particular, we observed that around $99\%$ of authors with the largest $1\%$ percentile values of $A_{50}$ belong to Physics \& Astronomy.
By excluding those authors from our analysis, we aim to mitigate this bias and gain a clearer understanding of metric $A_{50}$ in a broader context.
Thus, we end up examining $1,322,652$ authors. Among them, $242,281$ have a value greater than zero for the $A_{50}$ metric, they have at least $1$ co-author with whom they have co-authored more than $50$ papers.

For this metric the $99^{th}$ percentile value is $7$, and the $95^{th}$ percentile is $2$. The median is equal to $0$ as well as the interquartile range. Here, we focus on the top $1\%$ of authors who possess an $A_{50}$ value exceeding $7$, as we place particular emphasis on exceptional cases where an author has collaborated with a diverse set of individuals more than $50$ times, providing the strongest indication for large-scale orchestration.
Table~\ref{tab:field_authors_perc} indicates the allocation of these $12,015$ authors across scientific disciplines. A strong predilection is shown for authors within Clinical Medicine, as they represent $73.29\%$ of these $12,015$ authors, as opposed to only $38.71\%$ among all authors. No other field has an enrichment among the $12,015$ authors.
Supplementary table~\ref{tab:sup_clinical_medicine_subfields} shows the breakdown of Clinical Medicine authors into subfields. As shown, allergy, arthritis \& rheumatism, cardiovascular system \& hematology, gastroenterology \& hepatology, general clinical medicine, geriatrics, and psychiatry subfields are 1.5-fold or more enriched among those with $A_{50}$ values exceeding the highest $1\%$ percentile. 

\begin{figure*}
     \centering
     \begin{subfigure}[b]{0.45\textwidth}
         \centering
         \includegraphics[width=\textwidth]{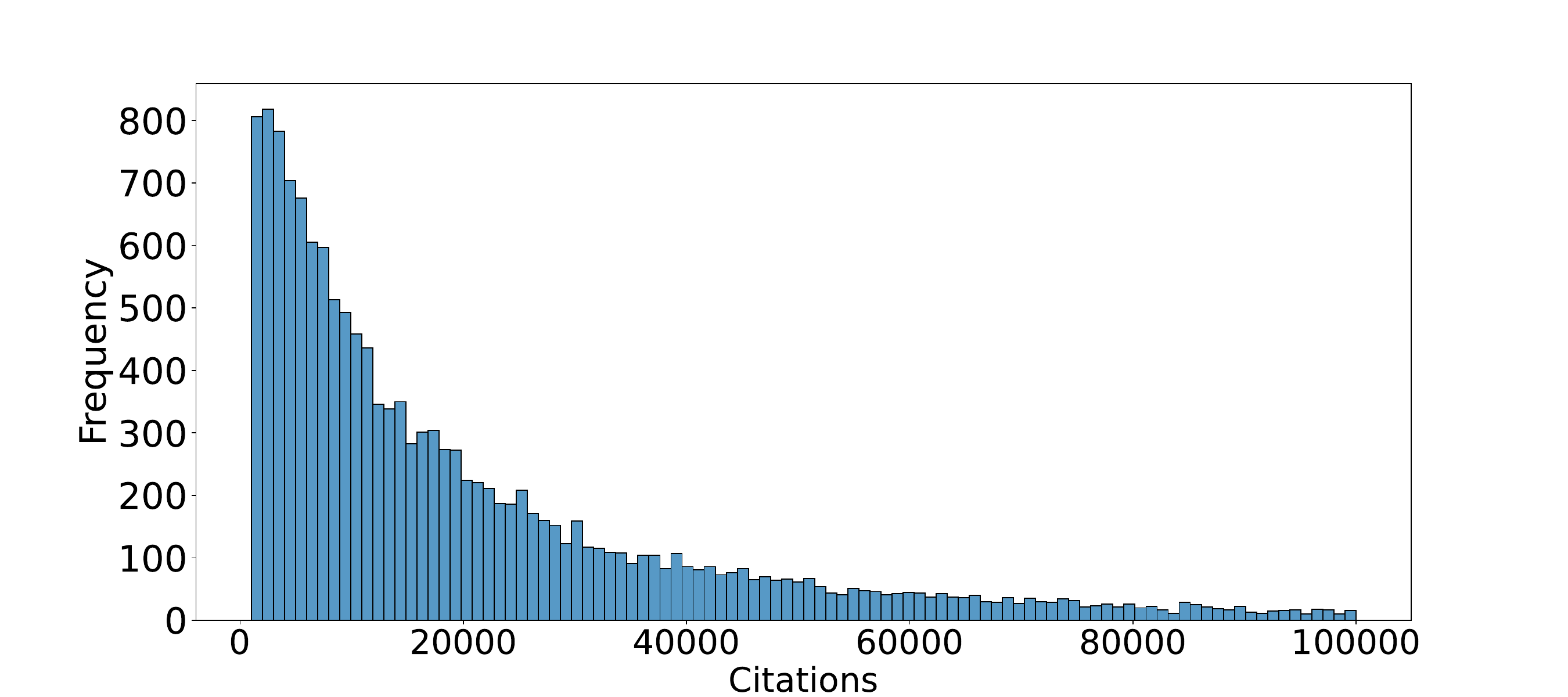}
        \caption{Distribution of the citations, considering only whoever author has an $A_{50}$ value greater than $7$ (authors with citations above $100,000$ are not shown).}
        \label{fig:elsevier_distr_cit_a6}
     \end{subfigure}
     \hfill
     \begin{subfigure}[b]{0.45\textwidth}
         \centering
         \includegraphics[width=\textwidth]{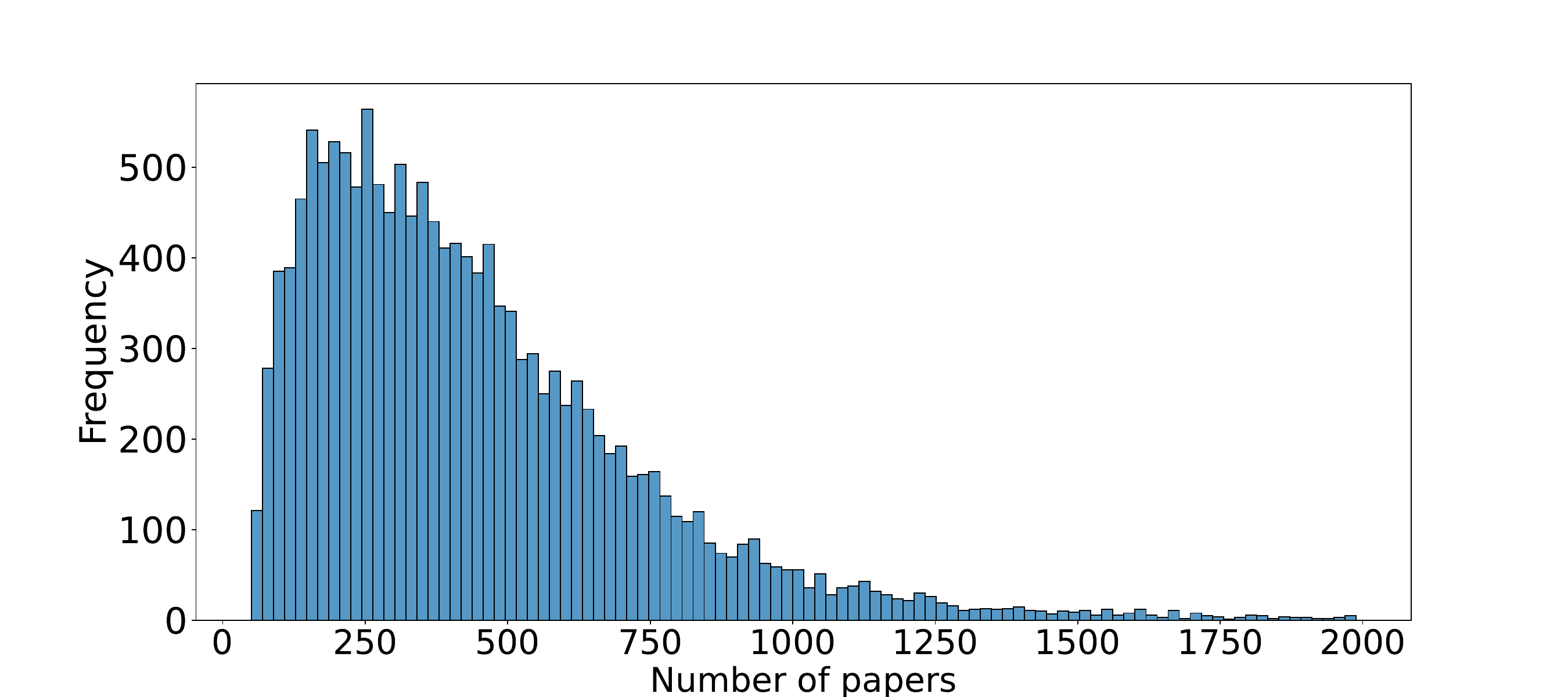}
        \caption{The distribution of the number of papers published by authors with an $A_{50}$ value on the upper 1\% (authors with more than $2,000$ papers are not shown).}
        \label{fig:elsevier_distr_numPapers_a6}
     \end{subfigure}
     \hfill
     \begin{subfigure}[b]{0.45\textwidth}
         \centering
         \includegraphics[width=\textwidth]{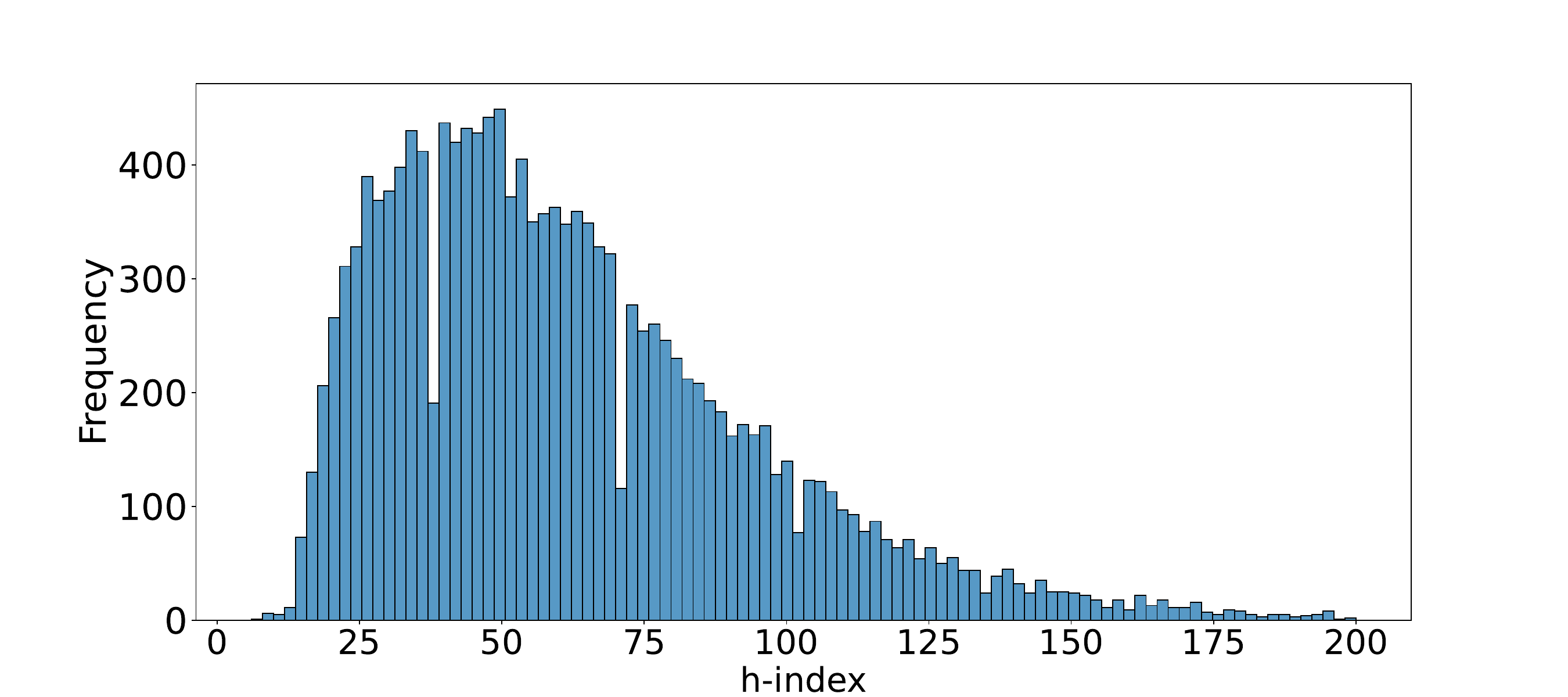}
        \caption{Distribution of the $h$-index of the authors that have an $A_{50}$ value greater than $7$ (authors with $h$-index above $200$ are not shown).
        }
        \label{fig:elsevier_distr_hindex_a6}
     \end{subfigure}
        \caption{The distribution of citations, number of papers and $h$-index of the authors that belong to the upper 1\% of metric $A_{50}$.}
        \label{fig:elsevier_cit_hindex_papers_a6}
\end{figure*}

Figure~\ref{fig:elsevier_cit_hindex_papers_a6} depicts the distribution of citations, the number of published papers, and the $h$-index, respectively, for this group of $12,015$ authors with the largest $A_{50}$ metric.
These are highly or even extremely productive authors (median $596$ papers, interquartile range $315$ to $1061$) and their $h$-indices are typically very high or even extraordinarily high (median $56$, interquartile range $38$ to $80$, $12.8\%$ of these authors with $h$-index>100).




\subsection{Co-existence of extreme values for the metrics}

We examined the co-existence of extreme values for the three metrics (lowest $1\%$ percentile for the first two, highest $1\%$ percentile for the third one) among all $1,322,652$ authors excluding Physics and Astronomy. As shown in Table~\ref{tab:coexistence}, there was strong co-existence for extremely low values of $C/h^2$ and $A_{50\%C}$ (odds ratio $6.4$, $95\%$ confidence interval $5.9-6.9$, $p<0.001$), modest coexistence for extremely low values of $A_{50\%C}$ and extremely high values of $A_{50}$ (odds ratio $1.5$, 95\% confidence interval $1.3-1.7$), and extremely low values of $C/h^2$ very rarely co-occurred with extremely high values of $A_{50}$ (odds ratio $0.09$, $95\%$ confidence interval $0.05-0.16, p<0.001$).

\begin{table*}[t]
    \begin{center}
    \scriptsize
    \renewcommand{\arraystretch}{1.2}
    \resizebox{\columnwidth}{!}{
    \begin{tabular}{|c|c|c|c|c|c|} 
        \cline{3-6} \multicolumn{2}{c|}{} & \multicolumn{2}{|c|}{Lowest percentile of $A_{50\%C}$} & \multicolumn{2}{|c|}{Highest percentile of $A_{50}$}\\
        \cline{3-6}
        \multicolumn{2}{c|}{} & YES & NO & YES & NO \\
        \hline
        \multirow{2}{*}{Lowest percentile of $C/h^2$} & YES & $659$ & $12,566$ & $11$ &  $13,214$\\
        & NO & $10,618$ & $1,298,809$ & $12,004$ & $1,297,423$ \\
        \hline
        \multirow{2}{*}{Lowest percentile of $A_{50\%C}$} & YES & & & $151$ & $11,126$\\
        & NO & & & $11,864$ & $1,299,511$\\
        \hline
    \end{tabular}}
    \end{center}
    \caption{Co-existence of small- and large-scale orchestration indicators}
    \label{tab:coexistence}
\end{table*}


 \subsection{Highly cited status among authors with orchestration indicators}


Of the authors in the lowest 1\% for the $C/h^2$ metric, 1,068/14,967 (7.1\%) are included in the 204,643 highly cited authors from the science-wide database. After excluding authors from the Physics \& Astronomy fields for the $A_{50}$ and $A_{50\%C}$ metrics, we found that 3,702/11,277 authors (32.8\%) in the upper 1\% of the $A_{50}$ metric are included in the database of the highly cited authors and 2,018/12,015 authors (16.8\%) in the lower 1\% of the $A_{50\%C}$ metric are also part of the database of highly cited authors.

\subsection{Retractions of publications among authors with orchestration indicators
}

A substantial number of authors with specific orchestration indicators (using the 1\% percentile) have had at least one retracted publication in Retraction Watch. The proportion of those with at least one retracted publication was very high (45.5\%, 5/11) for those with both $C/h^2$ and $A_{50}$. It was 5\% (5/100) for those with both $C/h^2$ and $A_{50\%C}$, for those  with both $A_{50\%C}$ and $A_{50}$, and for those with only extreme $C/h^2$. Conversely, it was 3\% (3/100) for those with only extreme $A_{50}$  and only 1\% (1/100) for those with only extreme $A_{50\%C}$.


\section{Discussion}
Using science-wide data from Scopus, we present three readily accessible metrics that may offer hints to orchestration patterns for single authors or groups of authors. This work may contribute to the ongoing efforts to enhance the assessment of research impact and recognize the contributions of authors within their respective fields~\cite{10.1371/journal.pbio.3002408, ioannidis2023quantitative}.  We found that authors with hints of orchestration based on these 3 metrics have enriched presence in specific scientific fields. Importantly, these authors almost always had high or even extremely high h-indices. These high $h$-indices could be misleading and would need to be interpreted in the light of the authorship and citation patterns to place them in a more proper context. 
A smaller set of these authors achieved high ranking even with a composite citation indicator that adjusts for co-authorship and authorship order patterns. Finally, we identified a substantial number of authors with extreme orchestration metrics, especially those involving extremely low $C/h^2$, who have already had retracted publications.

Authors at the extreme tails of these metrics may or may not engage in individual or group practices that lead to inflated authorship contributions and inflated perceived impact.  Moreover, most authors who do occasionally engage in practices that lead to inflated authorship contributions~\cite{masic2021inflated} or inflated perceived impact would not have such extreme values in these metrics. We propose to use these metrics with a very stringent threshold, such as 1\% percentile, where chance is less likely to generate such extremes in the absence of an organized plan or group arrangement regarding the allocation of authorship and/or citations.

However, any claims of ethical wrong-doing should be withheld in the absence of additional evidence. Authors with hints of extreme orchestration would need to be examined carefully as to the nature of their collaborations, teamwork arrangements and standards in their environment, and citation practices in the networks of scientists who cite them. Many of these scientists may have perfectly legitimate standards for their field, \eg massive co-authorship in their teamwork may be the widely accepted norm, as in the case of nuclear and particle physics work. Stable collaborations and teamwork are largely desirable in many fields. Nevertheless, even in these cases, the extreme nature of orchestration should be noted and the citation metrics of these authors should be properly adjusted (\eg for co-authorship~\cite{ioannidis2008measuring} and interpreted. This is essential in order to avoid unfair comparisons against other scientists in the same or neighboring fields who have different work and publication arrangements. Otherwise, comparisons would be grossly unfair.

We observed that the two metrics of small-scale orchestration, low $C/h^2$ and low $A_{50\%C}$, tended to coexist far more common than chance, which may reinforce the possibility of citation gaming for some authors.
Conversely, low $C/h^2$ and high \textit{$A_{50}$} rarely co-existed, but a large share of authors with these co-existing metrics already had retracted publications identified in Retraction Watch. Retraction rates were substantive (5\%) also for other authors with low $C/h^2$. This particular indicator has also been linked to spuriously ultra-precocious citation impact (reaching the status of highly cited researcher within a few years of starting to publish) as well as retraction risk in another empirical evaluation~\cite{Ioannidis2024.10.14.618366, soliman2025precocious}. Retraction data should be examined with caution given that probably most flawed or even outright fraudulent papers do not get retracted~\cite{oransky2022retractions}. As a result, only 0.7\% of authors with at least 5 full papers across science have had one retraction due to reasons other than publisher/journal error.
Regardless, citation orchestration may not be a phenomenon where classic forms of misconduct or flawed data are very common. Nevertheless, some organizations and initiatives, \eg the Swiss national code of conduct for researchers, are currently recognizing excessive self-citation practices as ethical violations~\cite{Matthews}. The presence of unjustified extreme self-citations or citation cartels should be assessed judiciously in each case. This requires an in-depth assessment of the nature, relevance, and justification of citations. Future work may also explore whether to what extent orchestration indicators may be linked to other unethical or spurious publication practices, such as paper mills~\cite{christopher2021raw}, hijacked journals~\cite{dadkhah2016hijacked,abalkina2021detecting}, hijacked citations~\cite{dadkhah2016types} and other subversive arrangements.

Retractions were rare in authors with large A50 alone, \ie those with the large-scale orchestration marker, despite the fact that these authors typically publish very high numbers of papers.
Given this high collaborative productivity, in authors with large-scale orchestration, sometimes astonishingly high citation counts and $h$-indices can be reached, but they are largely meaningless. These authors should be assessed for the nature of their contribution to the massively collaborative teamwork where they participate. 
Here we used an arbitrary level of co-authoring 50 papers with others, which may work well for medical fields, but lower numbers may need to be considered in the future in fields with less collaborative experience and lower levels of overall productivity.
Several unethical behaviors may co-exist in some cases of large-scale orchestration, \eg scientists may be placed as authors with gift authorship~\cite{smith1994gift,teixeira2016multiple}. This is a particular problem with people who acquire power as chairs/heads and exhibit a massive acceleration of their research productivity (at a time when administrative duties should probably have cut their productivity).
Moreover, while teamwork is a commendable research practice, in principle, inflation of authorship within teams may signal salami-slicing, and a poor sense of accountability and may reflect badly on the entire team/consortium.

Some limitations of our work should be discussed. First, we only focused on authors with at least 1,000 citations. Orchestration may also affect scientists who have fewer citations. Scientists with fewer than 1,000 citations are the vast majority of the scientific workforce. However, their influence in the scientific literature is limited compared with the more influential sample that we examined. Moreover, it is common for self-citations to be high in the early career of scientists when they have limited citations and they try to establish themselves with their work~\cite{SEEBER2019478}.  
This would possibly frequently raise signals of small-scale orchestration in early career authors that have a very different meaning compared with established authors with many citations. 
Second, inaccuracies in Scopus may affect the calculation of these metrics for these scientists. Inaccuracies are overall low across Scopus, but in the case of an in-depth assessment of the behavior of specific authors suspected of unethical orchestration practices, it is important to verify at the first step that their publication and citation profile is accurate, so as to avoid mischaracterizations. Third, we did not employ here more complex methods of community detection, centrality analysis, and network visualization~\cite{jolly2020unsupervised,fister2016toward}, but these methods may also be utilized when unethical orchestration is suspected and in-depth evaluation of specific author networks is needed. 

Overall, our work reveals that many authors with high or even exceptionally high traditional citation metrics such as the h-index have strong hints of orchestration practices. This distortion should be taken into account in using citation metrics for evaluation and reward purposes.

\bibliographystyle{apalike}
\bibliography{main}

\newpage
\appendix
\section{Supplementary material}\label{sec:sup}

\begin{table}[h!]
    \centering
    \resizebox{1\textwidth}{!}{
    \begin{tabular}{l|p{2.0cm}|p{2.0cm}|p{2.0cm}|p{2.0cm}}
        Subfield & Number of eligible authors & Percentage of eligible authors & Number of authors that belong to highest 1\% percentile & Percentage over the authors of highest 1\% percentile \\
        \hline
        allergy & $2,922$ & $0.50\%$ & $93$ & $1.06\%$ \\
        anesthesiology & $4,562$ & $0.79\%$ & $16$ & $0.18\%$ \\
        arthritis \& rheumatology & $6,614$ & $1.14\%$ & $168$ & $1.91\%$ \\
        cardiovascular system \& hematology & $61,481$ & $10.61\%$ & $1,462$ & $16.60\%$ \\
        dentistry & $8,488$ & $1.47\%$ & $22$ & $0.25\%$ \\
        dermatology \& venereal diseases & $7,218$ & $1.25\%$ & $77$ & $0.87\%$ \\
        emergency \& critical care medicine & $4,602$ & $0.79\%$ & $35$ & $0.39\%$ \\
        endocrinology \& metabolism & $25,727$ & $4.44\%$ & $373$ & $4.24\%$ \\
        environmental \& occupational health & $1,289$ & $0.22\%$ & $9$ & $0.10\%$ \\
        gastroenterology \& hepatology & $18,708$ & $3.23\%$ & $500$ & $5.68\%$ \\
        general \& internal medicine & $43,032$ & $7.43\%$ & $448$ & $5.09\%$ \\
        general clinical medicine & $287$ & $0.05\%$ & $9$ & $0.10\%$ \\
        geriatrics & $937$ & $0.16\%$ & $23$ & $0.26\%$ \\
        immunology & $46,471$ & $8.02\%$ & $470$ & $5.34\%$ \\
        legal \& forensic medicine & $601$ & $0.10\%$ & $5$ & $0.06\%$ \\
        neurology \& neurosurgery & $92,605$ & $15.98\%$ & $1,096$ & $12.45\%$ \\
        nuclear medicine \& medical imaging & $15,719$ & $2.71\%$ & $143$ & $1.62\%$ \\
        obstetrics \& reproductive medicine & $14,860$ & $2.56\%$ & $139$ & $1.58\%$ \\
        oncology \& carcinogenesis & $95,554$ & $16.49\%$ & $2,041$ & $23.18\%$ \\
        ophthalmology \& optometry & $10,206$ & $1.76\%$ & $65$ & $0.74\%$ \\
        orthopedics & $12,512$ & $2.16\%$ & $193$ & $2.19\%$ \\
        otorhinolaryngology & $4,199$ & $0.72\%$ & $19$ & $0.22\%$ \\
        pathology & $1,130$ & $0.19\%$ & $4$ & $0.05\%$ \\
        pediatrics & $9,012$ & $1.56\%$ & $56$ & $0.64\%$ \\
        pharmacology \& pharmacy & $21,038$ & $3.63\%$ & $64$ & $0.73\%$ \\
        psychiatry & $17,236$ & $2.97\%$ & $481$ & $5.46\%$ \\
        respiratory system & $11,667$ & $2.01\%$ & $179$ & $2.03\%$ \\
        sport sciences & $4,521$ & $0.78\%$ & $25$ & $0.28\%$ \\
        surgery & $18,604$ & $3.21\%$ & $303$ & $3.44\%$ \\
        tropical medicine & $3,024$ & $0.52\%$ & $25$ & $0.28\%$ \\
        urology \& nephrology & $14,419$ & $2.49\%$ & $263$ & $2.99\%$ \\
    \end{tabular}
    }
    \caption{The break-down of Clinical Medicine authors, who belong to the highest 1\% of metric $A_{50}$, into subfields }
    \label{tab:sup_clinical_medicine_subfields}
\end{table}

\end{document}